%% file: An_Efficient_Approach_for_Large-Scale_Video_Retrieval_by_Image_Queries.tex
\documentclass[times,twocolumn,final,authoryear]{elsarticle}

\usepackage{prletters}
\usepackage{framed,multirow}

\usepackage{amssymb}
\usepackage{latexsym}

\usepackage{url}
\usepackage{xcolor}

\usepackage{graphicx}
\usepackage{epstopdf}
\usepackage{geometry}

\usepackage{graphicx,amstext,amsmath}
\usepackage[justification=centering]{caption}

\usepackage{epsfig}
\usepackage{subfigure}
\usepackage{amsmath}
\usepackage{algorithmic}
\usepackage{graphics}
\usepackage{alltt}
\usepackage{float}
\usepackage{array}
\usepackage{xspace}
\usepackage{footmisc}
\usepackage{comment}
\usepackage{algorithm}
\usepackage{algorithmic}
\usepackage[misc]{ifsym}

\usepackage{multirow}
\usepackage{colortbl}

\definecolor{newcolor}{rgb}{.8,.349,.1}

\journal{Pattern Recognition Letters}

\begin{document}

\begin{frontmatter}

\title{CNN-VWII: An Efficient Approach for Large-Scale Video Retrieval by Image Queries}

\author[1]{Chengyuan \snm{Zhang}}

\author[1]{Yunwu \snm{Lin}}

\author[1]{Lei \snm{Zhu}}
\author[1]{Anfeng \snm{Liu}\corref{cor1}}
\cortext[cor1]{Corresponding author:
  Tel.: +86-137-2386-0232;}
\ead{afengliu@mail.csu.edu.cn}

\author[1]{Zuping \snm{Zhang}}
\author[1]{Fang \snm{Huang}}

\address[1]{ School of Information Science and Engineering, Central South University, Changsha 410083, China}

\begin{abstract}
This paper aims to solve the problem of large-scale video retrieval by a query image. Firstly, we define the problem of top-$k$ image to video query. Then, we combine the merits of convolutional neural networks(CNN for short) and Bag of Visual Word(BoVW for short) module to design a model for video frames information extraction and representation. In order to meet the requirements of large-scale video retrieval, we proposed a visual weighted inverted index(VWII for short) and related algorithm to improve the efficiency and accuracy of retrieval process. Comprehensive experiments show that our proposed technique achieves substantial improvements (up to an order
of magnitude speed up) over the state-of-the-art techniques with similar accuracy.
\end{abstract}


\end{frontmatter}

\input{Introduction}

\input{RelateWork}
\input{ProblemDefinition}

\input{Model}
\input{Index}

\input{Experiment}

\input{Conclution}

\input{Acknowledgement}

\bibliographystyle{model2-names}
\bibliography{refs}

\end{document}

%% file: Introduction.tex
\section{Introduction}
\label{intro}
With advances in Internet technologies and the proliferation of smart phone, digital cameras, storage devices, there are a rapidly growing amount of video-related data collected in many applications, such as video search, recommendation, sharing, broadcasting and advertising websites. As a result, in recent years various visual search applications have emerged such as image to video retrieval, video to image retrieval and video to video retrieval.

In the paper, we investigate the problem of retrieving top-$k$ most relevant videos for a large video database by using an image as the query; that is, given a set of video, a query image, we aim to retrieve the $k$ most relevant videos each of which contains frames which are similar to query image. For example, users can find the video of a movie by a snap without the name on the Internet. Another application is that a user can search out a lecture video online by a slide.

Considerable effort has been invested by the research community on image to video retrieval problem.
The simplest solution to this problem is to consider each frame in video as an individual image. Thus, the problem of detecting whether an image is contained by an video is converted to the problem that whether an image is similar to another image. Although this solution can achieve high accuracy, a large number of unnecessary image to image comparison are carried out, which is time consuming and prohibitive for large scale video databases.

Most recently, replacing image by video clip has shown great success in image to video retrieval in \cite{DBLP:journals/tcsv/AraujoG18}. Thus, the search space reduces from all frame in video database to a small group of video clips, which significantly improve the scalability of the retrieval process. Specifically, the query image's descriptor is first compared to the bloom filter index that contains video clip level information.
Then, the frames belonging to the most promising video clips are examined one by one to calculate the similarities with the query image. Although bloom filter technique can effectively improve the retrieval efficiency, it still face the following challenges. Firstly, comparing video clip level bloom filter one by one is still unacceptable for large video databases. Secondly, the false positive probability of bloom filter will increase, as the number of frames in video clip increases. Thirdly, bloom filter of video clips level cannot well preserve the aggregate information from the frame level.

This work aims to improve the performance of video retrieval by a query image in a large-scale video database. First, we introduce the definition of image to video query, and devise the visual similarity function to measure the visual similarity between query image and frame of video. Base on these notions, we propose a novel solution to improve the performance of video retrieval by a query image, which is a combination of CNNs, bag-of-visual-words and inverted index technique. Specifically, We utilize convolutional neural network to construct the visual feature extractor in our solution. All the frames of videos in database are fed into this extractor and the features are extracted, which are used to generate the BoVW model of visual representation. The frames of videos are grouped by $K$-means algorithm to reduce the size of index significantly. Besides, a novel indexing structure named VWII and an efficient search algorithm are developed, which can be used to boost the search performance.

\noindent\textbf{Contributions}. The main contributions of this work can be summarized as follows:

\begin{itemize}
\item We introduce the formal definition of image to video top-$k$ query problem, and present the similarity function to measure the visual similarity between query image and frame of video.
\item We propose a novel solution to improve the efficiency and precision of the query, which is based on convolutional neural networks, bag-of-visual-words and visual weighted inverted index.
\item We devise an efficient search method named VWII Search algorithm based visual weighted inverted index technique. This algorithm can boost the performance of video search in a large-scale database.
\item We have conducted extensive performance evaluation on two video dataset. Experimental results demonstrate that our approach outperforms the state-of-the-art method.
\end{itemize}

\noindent\textbf{Roadmap}. In the remainder of this paper, Section~\ref{relate-work} introduce the related works about this paper. In Section~\ref{problem-definition}, we propose the definition of image to video top-$k$ query and give the visual similarity function. In Section~\ref{our-model}, we introduce our model to represent video and query image and measure the similarity between image and video. In Section~\ref{index}, we devise a novel indexing structure named VWII and present an efficient search algorithm. Section~\ref{perf-evo} presents the the experiment results on two video datasets. In Section~\ref{conclution}, we conclude the paper.

%% file: RelateWork.tex
\section{Related Work}
\label{relate-work}

In this section, we introduce the related works of large-scale video retrieval, which includes video to video problem, image to video problem and video to image problem.

\noindent\textbf{Video to Video Problem}. Video to video problem if one of the most common search problem in the area of video retrieval, such as~\cite{DBLP:journals/pr/WuWGL18}. It aims at finding the most similar videos in the aspect of visual contents from the database. Lots of visual representation and search techniques have been proposed by researches in recent years. Bag-of-words (BoW for short) is one of traditional visual representation methods for image and video representation, which is proposed by~\cite{DBLP:conf/iccv/SivicZ03} who used this model to index videos throughout a movie database. ~\cite{DBLP:journals/mms/UlutasUUN18} presented a novel approach to solve the problem of frame duplication detection in a video. They utilized BoW model to generate visual words and build a dictionary from Scale Independent Feature Transform (SIFT for short) keypoints of frames in video. They adopted Hierarchical $k$-means (HKM) to generate a large vocabulary tree for quantization, and represent each video clip and query topic by a BoW model. ~\cite{DBLP:conf/cikm/WangSE12} integrated the spatial-temporal information with BoW model to improve computational efficiency significantly. In specific, they model the pair-wise spatial-temporal correlations by a Gaussian distribution and devised a novel similarity measurement to emphasizes the discriminative visual words about the query.~\cite{DBLP:conf/sigir/JiangN08} introduced a novel method extending from BoW model to reduce the effect of visual word correlation. In their approach, a visual ontology was generated to model the $is-a$ relationship between visual words, in which visual relatedness was defined strictly and then incorporated into BoW model.

Due to the great advantages of deep learning technology in multimedia retrieval, such as~\cite{DBLP:journals/pr/WuWLG18}, more and more researchers utilize deep neural networks such as CNN for the task of video retrieval, such as~\cite{DBLP:journals/corr/abs-1808-01911}.~\cite{DBLP:journals/corr/PodlesnayaP16} utilized CNN to extract visual features of videos, which serve as universal signature for retrieval tasks. For the problem of near-duplicate
video retrieval which is one of the types of video to video problems, ~\cite{DBLP:conf/iccvw/Kordopatis-Zilos17} a novel scheme by using CNN features from intermediate layers to create discerning global video representations integrating with a deep metric learning framework.~\cite{DBLP:conf/dcc/LouBLW0CDHK017} used a novel deep-learning features in CDVA evaluation framework for video retrieval problem. Specifically, they devised a Nested Invariance Pooling (NIP for short) method to generate compact and robust CNN descriptors.~\cite{DBLP:conf/mm/GuMY16} combined CNN and Long-short Term Memory (LSTM for short) Networks to implement Supervised Recurrent Hashing (SRH for short) to overcome the challenge of large-scale video retrieval.

\noindent\textbf{Image to Video Problem}. For image to video problem, many previous researches were inspired by the solutions for traditional image retrieval tasks.~\cite{DBLP:journals/ijcv/SivicSZ06} presented a novel method which automatically generates object representations for the task of image query that is to return the objects of interest in video shots. For the task of relevant video segments searching, ~\cite{DBLP:conf/mir/ZhuS12,TIP15YangWang,YLNN2018} formulated this problem by a large vocabulary quantization based Bag-of-Words framework.~\cite{DBLP:conf/icip/AraujoCAG15} proposed to solve the problem of large-scale video retrieval by image query based on binarized Fisher Vectors. They presented an asymmetric comparison scheme to improve the mean average precision. Besides, Several shot-based aggregation techniques was developed by them to achieve retrieval performance with a 3X speed-up. For the task of face video retrieval by image query,~\cite{DBLP:conf/cvpr/LiWHSC15} presented Hashing across Euclidean space and Riemannian manifold based on a unified framework to integrate the two heterogeneous spaces into a common discriminant Hamming space. Then the intra-space and inter-space Hamming distances are optimized in a maxmargin framework, which is to generate hash functions.~\cite{DBLP:conf/aaai/ZhuJWWZZ17} introduced a joint feature projection matrix and heterogeneous dictionary pair learning (PHDL for short) approach to solve the problem of image to video person re-identification, which is an important technique in video surveillance. This approach learns an intra-video projection matrix and a pair of heterogeneous image and video dictionaries, in which the heterogeneous visual features can be turned into coding coefficients. These coding coefficients can be used to implement visual matching.~\cite{DBLP:conf/nips/TangR0K12,DBLP:journals/corr/abs-1804-11013,DBLP:journals/tnn/WangZWLZ17,NNLS2018,DBLP:journals/tip/WangLWZ17,DBLP:conf/mm/WangLWZ15,DBLP:conf/ijcai/WangZWLFP16} investigated the problem of adapting detectors from image to video and introduced a novel approach to overcome this challenge. They classify tracks to discover examples in the unlabeled videos to leverage temporal continuity. In addition, they designed a new self-paced domain adaptation algorithm to iteratively adapt from source domain to target domain.~\cite{DBLP:journals/tcsv/AraujoG18,TC2018,DBLP:journals/cviu/WuWGHL18} developed an asymmetric comparison method by using Fisher vectors for large-scale video retrieval by image query. A novel video descriptors which can be compared directly with image descriptors was devised by them. 

%% file: ProblemDefinition.tex
\section{Problem Definition}
\label{problem-definition}

In this section, we introduce the definition of video retrieval by image query and related notions. Table ~\ref{tab:notation} summarizes the mathematical notations used throughout this paper to facilitate the discussion of our work.

\begin{table}
	\centering
    \small
	\begin{tabular}{|p{0.20\columnwidth}| p{0.70\columnwidth} |}
		\hline
		\textbf{Notation} & \textbf{Definition} \\ \hline\hline
		~$\mathcal{V}$                                   & A given database of videos      \\ \hline
        ~$|\mathcal{V}|$                                 & The number of videos in $\mathcal{O}$     \\ \hline
        ~$V$                                             & A video    \\ \hline
	  	~$f$                                             & A frame in a video        \\ \hline
        ~$V.\mathcal{F}$                                 & The frame set of video $V$ \\ \hline
        ~$I_q$                                           & A query image      \\ \hline
        ~$I.\Omega$                                      & The visual word set of image $I$ \\ \hline
        ~$\mathcal{Q}_{I2V}$                             & A image to video query       \\ \hline
        ~$\mathcal{R}_Q$                                 & The result set of $\mathcal{Q}_{I2V}$      \\ \hline
        ~$Sim(I_q,V)$                                    & The similarity between $I_q$ and $V$      \\ \hline
		~$\omega$                                        & A visual word    \\ \hline
        ~$|I_q|$                                         & The number of visual words in image $I_q$   \\ \hline
		~$|f|$                                           & The number of visual words in frame $f$        \\ \hline
        ~$VisSim(I,f)$                                   & The visual similarity between $I$ and $f$            \\ \hline
        ~$W(\omega)$                                     & The weight of visual word $\omega$             \\ \hline
        ~$V.\mathcal{C}$                                 & The frame cluster set of video $V$       \\ \hline
        ~$\mathcal{D}$                                   & The visual dictionary    \\ \hline
	\end{tabular}
    \caption{The summary of notations} \label{tab:notation}	
\end{table}

\noindent\textbf{Top-$k$ Image to Video Query}. Let $\mathcal{V} = \{V_1,V_2,...,V_{|\mathcal{V}|}\}$ be a video database which contains $|\mathcal{V}|$ videos, where $V.\mathcal{F} = \{f_1,f_2,...,f_{|V.\mathcal{F}|}\}$ is a video containing $|V.\mathcal{F}|$ successive frames. Let $I_q$ be a query image and $k$ be a positive integer, a top-$k$ image to video query is denoted as $Q^k_{I2V}$ which aims to return a set containing $k$ videos $\mathcal{R}_Q$ in which each video contains the visual content of $I_q$. Formally, we denote $\mathcal{R}_Q$ as,
\begin{equation*}
\mathcal{R}_Q=\{V|Sim(I_q,V) > Sim(I_q,V'), \forall V \in \mathcal{V},V' \in \mathcal{V} \setminus \{V\}\},
\end{equation*}
\begin{equation*}
|\mathcal{R}_Q| = k
\end{equation*}
where $Sim(I_q,V)$ represents a similarity function which is to measure the visual similarity between an image $I_q$ and a video $V$. The more similar $I_q$ and $V$ in the aspect of visual content, the higher the value of $Sim(I_q,V)$ is.

In this work, we use the traditional visual representation technique named Bag-of-Visual-Word (BoVW for short) to represent the visual content of image and video. Specifically, for the query image $I_q$, it is encoded by a bag-of-words vector with several visual words, denoted as $I_q.\Omega=\{\omega_1,\omega_2,...,\omega_{|I_q.\Omega|}\}$, $|I_q.\Omega|$ represent the number of visual works. In the same way, for a frame $f$ in a video, it can be denoted as $f=\{\omega_1,\omega_2,...,\omega_{|f|}\}$.

\noindent\textbf{Visual Similarity between image and frame}. Given a image $I$, $I.\Omega=\{\omega^I_1,\omega^I_2,...,\omega^I_n\}$ and a video $V = \{f_1,f_2,...,f_n\}$, $\forall f \in V$, the visual similarity between $I$ and $f$ is measured by the following similarity function:
\begin{equation}\label{equ:vis-sim-i-f}
VisSim(I,f) = \frac{\sum_{\omega \in I.\Omega \cap \omega \in f}^{}W(\omega)}{\sum_{\omega \in I.\Omega \cup \omega \in f}^{}W(\omega)}
\end{equation}
where $W(\omega)$ is the weight of the visual word $\omega$. Inspired by the solution in text mining, in this work we measure the weight of visual word by term frequency-inverse document frequency $TF\mbox{-}IDF$, shown as follows:
\begin{equation}\label{equ:tf-idf}
W(\omega) = F_T(\omega)*(log\frac{N+1}{N(\omega)+1}+1)
\end{equation}
where $F_T(\omega)$ is the word frequency of $\omega$ in the current visual word dictionary, $N$ is the total number of frames in the database, $N(\omega)$ represents the total number of frames containing $\omega$ in the database.

%% file: Model.tex
\section{Our Model}
\label{our-model}
In this section, we propose our model for visual feature extraction and representation based on convolutional neural networks and bag-of-visual-word model. At first we present the overall of this model. Then, we introduce the visual feature extraction by CNNs and present how to represent video based on a visual dictionary construct from video database.

\begin{figure*}[thb]
\newskip\subfigtoppskip \subfigtopskip = -0.1cm
\centering
\includegraphics[width=1.0\linewidth]{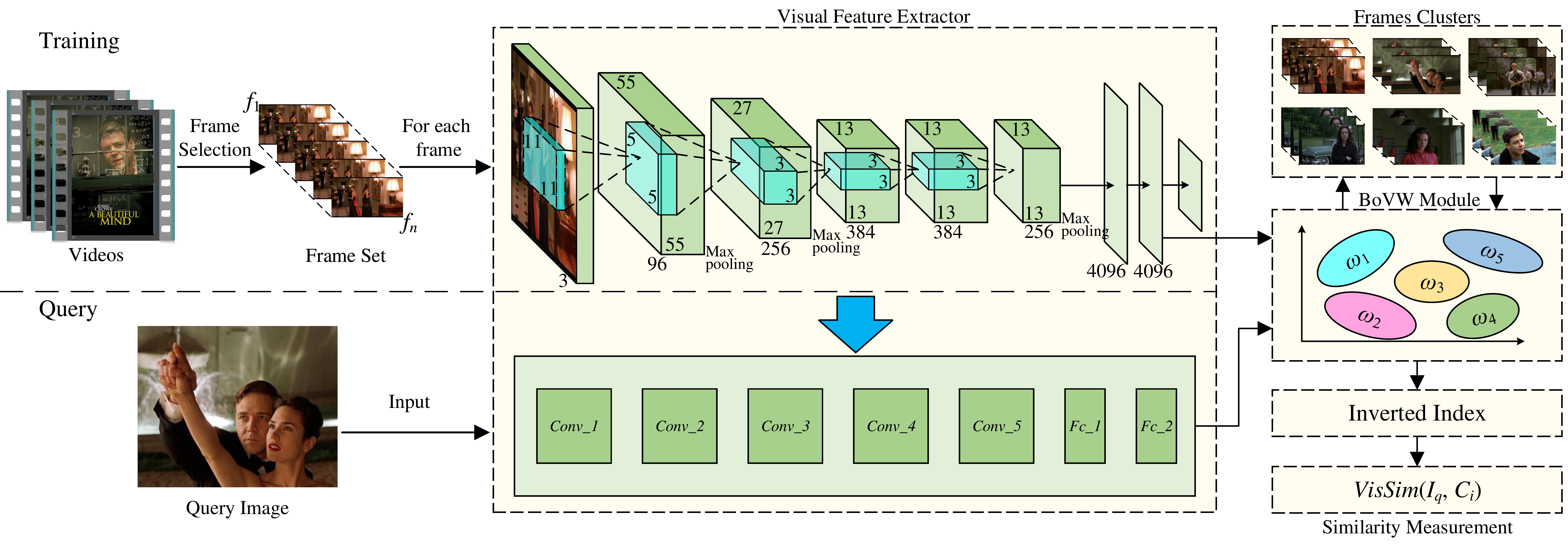}
\vspace{-1mm}
\caption{\small  The overview of our model }
\label{fig:model}
\end{figure*}

\subsection{Overview of Our Model}

Fig.~\ref{fig:model} demonstrates the overview of our model for visual feature extraction and video representation by utilizing CNNs and BoVW model. It aims to extract the visual features of videos from a database and then construct a visual dictionary by using clustering method. For each video, we represent it as a set of blocks in which each block consists of several frames which are similar to each other. More concretely, Given a video database, we select each of the video $V$ and construct a set of frames $V.\mathcal{F}=\{f_1,f_2,...,f_{|V.\mathcal{F}|}\}$ by using sampling at fixed-time intervals technique to select frames from the video. After video frames selection, all the frames in $\{V_1.\mathcal{F},V_2.\mathcal{F},...,V_n.\mathcal{F}\}$ are fed into the feature extractor implementing by convolutional neural networks to extract the visual features. For each video, the CNNs extracts the low-level visual features and construct a feature vector and then put all the vectors into the BoVW module to training the visual dictionary $\mathcal{D}$ by using clustering methods. Based on the visual dictionary $\mathcal{D}$, the frames in each video are clustered into several frame clusters according to the visual similarity and then the video $V$ is represented by a series of frame clusters, denoted as $V.\mathcal{C} = \{c_1,c_2,...,c_{|V.\mathcal{C}|}\}$, and $\forall c_i \in V.\mathcal{C}$, $c_i = \{f^{(i)}_1,f^{(i)}_2,...,f^{(i)}_{|c_i|}\}$.

For the task of query image $I_q$ representation, we treat it as a frame of the video. That is, we directly feed it into the feature extractor and the use the similar technique mentioned above to generate the visual word vector $I_q.\Omega = \{\omega_1,\omega_2,...,\omega_{|I_q|}\}$.

\subsection{The Feature Extractor}
We utilize convolutional neural networks to implement the visual feature extractor. CNNs is one of the most significant neural networks, it has become a hotspot in the field of image classification, computer vision, speech analysis, etc..

In this work, we use the AlexNet architecture to construct our model, which is an important deep neural network model for the task of image recognition introduced by~\cite{DBLP:conf/nips/KrizhevskySH12}. AlexNet has 650,000 neurons and 60 million parameters, which consists of five convolutional layers, some of which are followed by max-pooling layers, and three fully-connected layers and a 1000-way softmax at last. The activation functions of neurons in this network are modeled by Rectified Linear Units (ReLUs). A 4096-dimensions feature vector can be generated from the first seven layers (the five convolutional layers and the first two fully-connected layers) of the network. For the input video $V_i$, this CNNs based feature extractor can be seen as a non-linear function $E_{cnn}(f_j)$, which is to return a visual feature vector $\Gamma_j$ from a input frame of a video $f_j \in V_i.\mathcal{F}$, i.e.,
\begin{equation*}
\Gamma_j = E_{cnn}(f_j), \forall f_j \in V_i.\mathcal{F}
\end{equation*}
and
\begin{equation*}
\Gamma_j = (\gamma^{(1)}_j,\gamma^{(2)}_j,...,\gamma^{(k)}_j,...,\gamma^{(4096)}_j)^T
\end{equation*}
where $\gamma^{(k)}_j$ represents the $k$-th-dimension visual feature of $f_j$. Apparently, the output of feature extractor for a video $V_i$ is a matrix $\textbf{M} = (\Gamma_1,\Gamma_2,...,\Gamma_{|V_i.\mathcal{F}|})$.

Each frame $f$ fed into the network is resized into $256 \times 256$. The number of neural units of the second-to-last fully-connected layer is 4096. The 4096-dimension feature vector of it is used as the input of the BoVW module.

\subsection{The BoVW Module}
Similar to the processing in the text similarity computation, the BoVW module aims at constructing a visual dictionary for the target video database based on the output of the feature extractor. We use K-means clustering method to implement the converting from visual features into visual words. As one of the most widely used unsupervised learning algorithm, K-means accepts an unlabeled data set and clusters the data into different groups. Given the visual feature vectors set $\{\Gamma_1,\Gamma_2,...,\Gamma_n\}$, K-means algorithm partition the sample set into $K$ clusters $\mathcal{G} = \{G_1,G_2,...,G_K\}$, the objective function can be defined as follows:
\begin{equation}
F_{K-means} = \mathop{\arg\min}_{\mathcal{G}=\{G_i\}^K_{i=1}}\sum_{i=1}^{K}\sum_{\Gamma_j}^{G_i} \lVert \Gamma_j-\mu_i \rVert ^2_2
\end{equation}
where $\mu_i$ is the mean vector of the cluster $G_i$, i.e., $\mu_i = \frac{1}{|G_i|}\sum_{\Gamma \in G_i}^{}\Gamma$.

\subsection{Frame Clusters Construction}
In order to speed up the retrieval performance, we improve the video representation based on frames and propose the novel representation named frame Clusters. Specifically, for a video $V.\mathcal{F} = \{f_1,f_2,...,f_{|V.\mathcal{F}|}\}$, we cluster these frames according to the visual similarity by using K-means algorithm. After the clustering, all the frames of $V$ are grouped into $K$ frame clusters, denoted as $V.\mathcal{C} = \{C_1,C_2,...,C_{|V.\mathcal{C}|}\}$, and $\forall C_i \in V.\mathcal{C}, C_i = \{f^i_1,f^i_2,...,f^i_{|C_i|}\}$. For each frame cluster $C_i$, we can construct the visual word set of it in the following manner:
\begin{equation}
C_i.\Omega = f^i_1.\Omega \cup f^i_2.\Omega \cup ... \cup f^i_{|C_i|}.\Omega
\end{equation}

In the processing of video retrieval, we only needs to compute the visual similarity between the query image and each frame cluster of each video. We set a threshold $\epsilon$ to measure the similarity. That is, $\forall V \in \mathcal{V}$, and $\forall C_i \in V.\mathcal{C}$, if $VisSim(I_q,C_i) \geq \epsilon$, then $I_q$ is similar to $V$. 

%% file: Index.tex
\section{Visual Weighted Inverted Index}
\label{index}

\begin{figure}[thb]
\newskip\subfigtoppskip \subfigtopskip = -0.1cm
\centering
\includegraphics[width=1.0\linewidth]{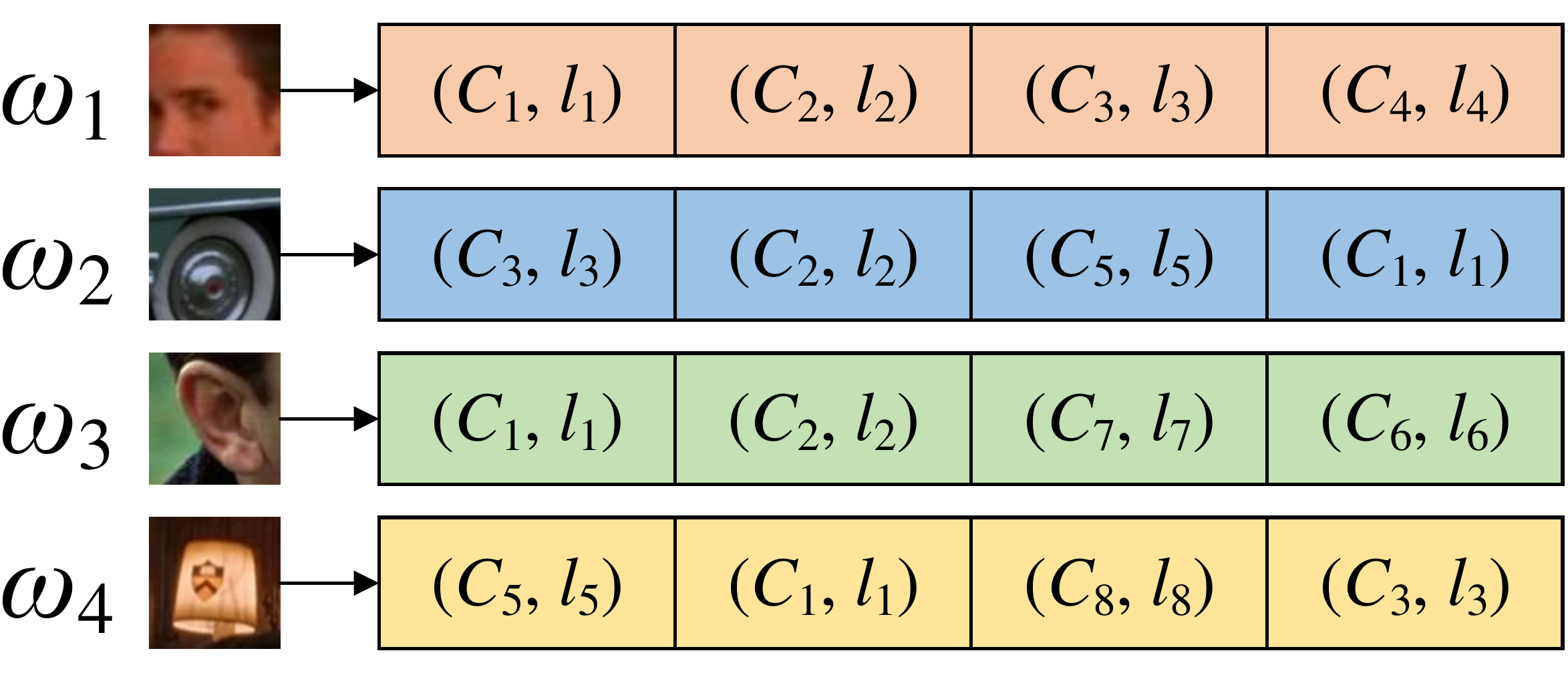}
\vspace{-1mm}
\caption{\small  VWII structure }
\label{fig:index}
\end{figure}

In order to improve the search performance in large-scale video database, we propose to construct an efficient index for videos and develop an algorithm based on it. In this section, we present a novel indexing structure named VWII, which is a combination of visual words of videos and inverted index. First we introduce the framework of our indexing structure and then propose our search algorithm.

\subsection{Our Index Framework: VWII}
We model the image to video query as a visual $\epsilon$-aggregation problem, which is inspired by the aggregation problem in the area of database~\cite{DBLP:conf/pods/FaginLN01,DBLP:conf/sigir/ZhangCT14}.

\noindent\textbf{Visual $\epsilon$-Aggregation Problem}. Consider a video frame set $\mathcal{F}$ where each frame $f=\{\omega_1,\omega_2,...,\omega_{|f|}\}$ has $|f|$ visual words. Let $F_a(\cdot)$ be a monotonic aggregation function, where $F_a(f)$ denote the overall visual score of frame $f$. Given a similarity threshold $\epsilon$, the visual $\epsilon$-aggregation problem aims to return a set of frames $f \in \mathcal{F}$ with the visual score is larger than or equal to $\epsilon$, i.e., return a set $\mathcal{R}_f \in \mathcal{F}$, in which $\forall f \in \mathcal{R}_f$ and $\hat{f} \in \mathcal{F} \setminus \mathcal{R}_f, F_a(f) \geq F_a(\hat{f})$.

It is suitable that define the image to video query problem as the visual $\epsilon$-aggregation problem. For a large-scale video database $\mathcal{V}$, frames can be selected by frame sampling techniques and a frame set $\mathcal{F}$ of them can be constructed. To put it in another word, the database $\mathcal{V}$ is modeled by $\mathcal{F}$ straightforward. Given a query image $I_q$, the similarity between $I_q$ and a frame $f$ can be measured by the visual similarity function $VisSim(I_q, f)$ and the similarity threshold $\epsilon$. For a frame $f_i \in \mathcal{F}$, if $VisSim(I_q, f_i) \geq \epsilon$, then $f_i$ can be added into $\mathcal{R}_f$. Therefore, when the search is terminated, we can easily to know that $\forall f \in \mathcal{R}_f$ and $\hat{f} \in \mathcal{F}-\mathcal{R}_f$, $F_a(f) \geq F_a(\hat{f})$.

By defining the image to video query problem as a visual $\epsilon$-aggregation problem, we can devise an indexing structure which is dependent on the inverted index rather than complex hybrid methods to solve this problem efficiently. Thus, we integrate visual words of frames and the inverted index technique to construct a novel index named \textbf{V}isual \textbf{W}ord \textbf{I}nvered \textbf{I}ndex (VWII for short) Fig.~\ref{fig:index} illustrates an example of VWII structure combining visual words and inverted indexing structure used in image to video query. Given a query image with $\alpha$ visual words, a list set containing $\alpha+1$ sorted lists $\mathcal{L}=\{L_1,L_2,...,L_{\alpha+1}\}$ is generated, and for each list $L_i \in \mathcal{L}$, it is sorted by a descending order based on the $TF\mbox{-}IDF$ score of the $i$-th visual word.

\subsection{Search Algorithm}
\begin{algorithm}
\begin{algorithmic}[1]
\footnotesize
\caption{\bf VWII Search Algorithm}
\label{alg:vwiisearch}

\INPUT  A query image $I_q$, a positive integer $k$, the number of query visual words $\beta$.
\OUTPUT A results set $\mathcal{R}_k$.

\STATE Initializing: $\mathcal{R}_k \leftarrow \emptyset$;
\STATE Initializing: $S_{min} \leftarrow 0$;
\STATE $B_k \leftarrow 1$;

\FOR{$i \leftarrow 1$;$i \leq m$;$i++$}
    \STATE $PTR[i] \leftarrow 0$;
\ENDFOR
\WHILE{$S_{min} \leq B_k$}
    \FOR{$i \leftarrow 1$;$i \leq m$;$i++$}
        \FOR{$j \leftarrow 0$;$j \leq \xi$;$j++$}
            \STATE $d \leftarrow L_t[i][PTR[i]]$;
            \STATE $ACC(FrmID,i,FrmID.score)$;
            \STATE $PTR[i] \leftarrow PTR[i]+1$;
        \ENDFOR
    \ENDFOR
    \FOR{each available frame $f \in \mathcal{R}_k$}
        \STATE $RACC(FrmID)$;
        \STATE $UpdateSet(\mathcal{R}_k,S_{min})$;
    \ENDFOR
    \STATE $B_k= \beta * B_V(i)$;
\ENDWHILE
\FOR{each $FrmID \in \mathcal{F}$}
    \IF{$FrmID \notin \mathcal{R}_k$ \&\& $FrmID$ is available}
        \STATE $RACC(FrmID)$;
        \STATE $Uptdate(\mathcal{R}_k,S_{min})$;
    \ENDIF
\ENDFOR
\RETURN $\mathcal{R}_k$;
\end{algorithmic}
\end{algorithm}
Base on VWII structure, we develop a efficient search algorithm named VWII Search to improve the search performance in large-scale video databases. In this subsection, we discuss this algorithm in details.

In VWII Search algorithm, we use a parameter $\xi$ to control the successive access. In each iteration, there are $\xi$ frame clusters in each list are accessed successively. For each accessed frame $f$, $B(f)$ is defined as an upper bound of the aggregated score of this cluster. A frame is considered as an available frame if the bound $B(f)$ is greater than the highest score that has been computed so far. When an iteration is to the end, the available frame with the maximum bound is chosen for random access to compute the aggregated score. Algorithm~\ref{alg:vwiisearch} shows the pseudo-code of VWII Search algorithm.

%% file: Experiment.tex
\section{Performance Evaluation}
\label{perf-evo}

\begin{figure*}
\newskip\subfigtoppskip \subfigtopskip = -0.1cm

\begin{minipage}[b]{0.49\linewidth}
\begin{center}
     \subfigure[\footnotesize{YouTube-8M}]{
     \includegraphics[width=0.48\linewidth]{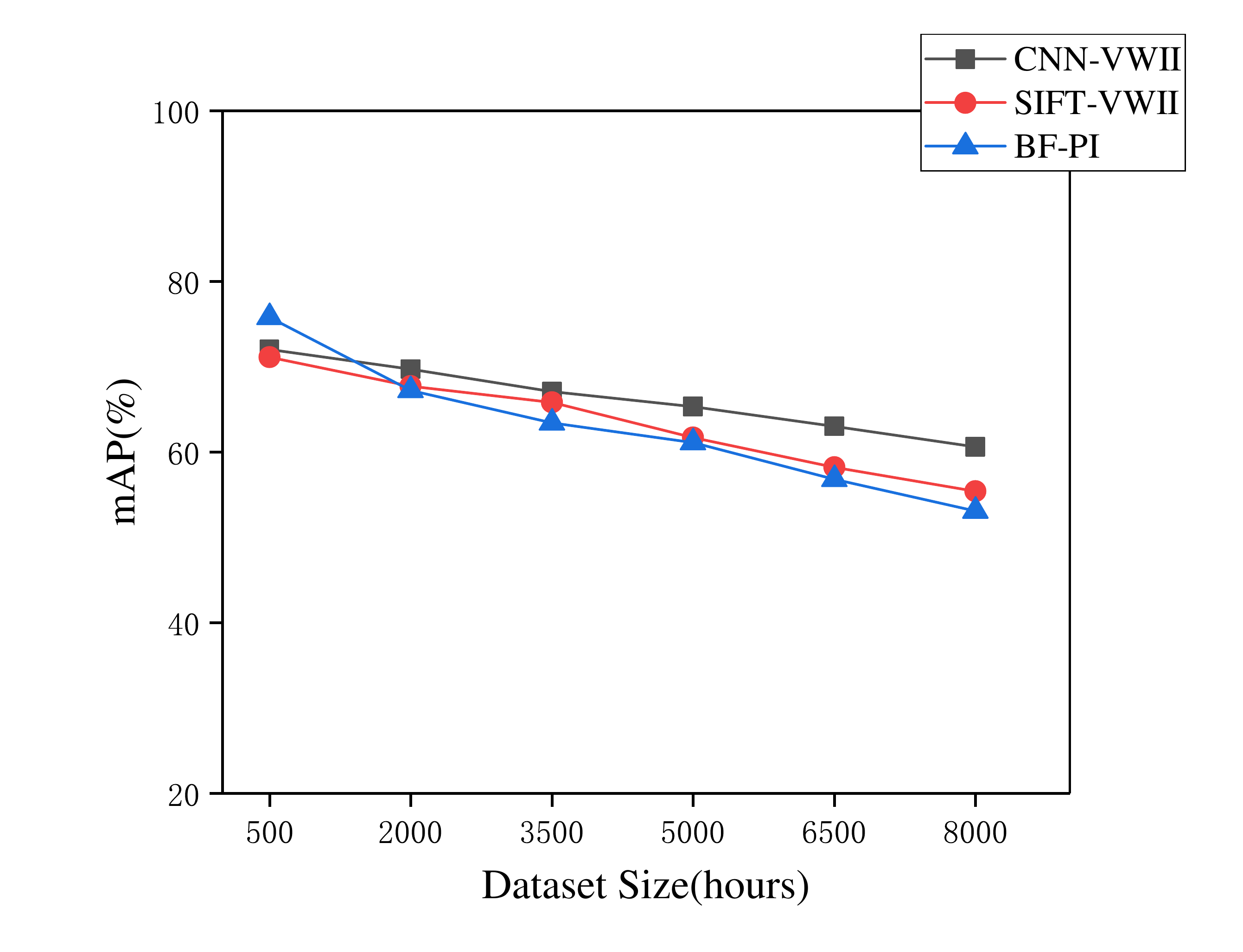}
     }
     \subfigure[\footnotesize{Sports-1M}]{
    \includegraphics[width=0.48\linewidth]{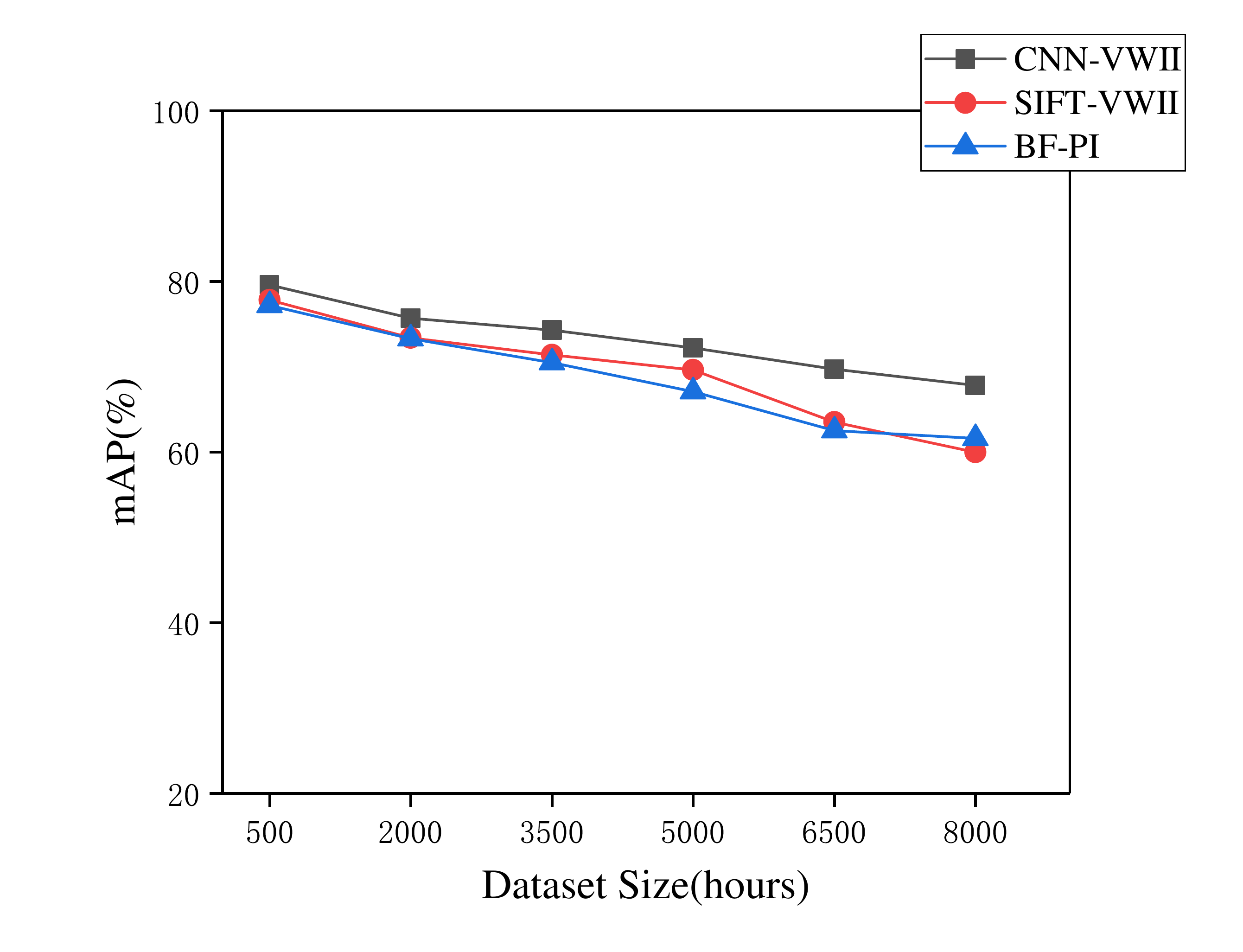}%
     }
     \captionsetup{justification=centering,font={footnotesize,sf}}
       \vspace{-0.3cm}
\caption{Precision evaluation on the size of dataset}
\label{fig:map-dataset-size}
\end{center}
\end{minipage}
\begin{minipage}[b]{0.49\linewidth}
\begin{center}
     \subfigure[\footnotesize{YouTube-8M}]{
     \includegraphics[width=0.48\linewidth]{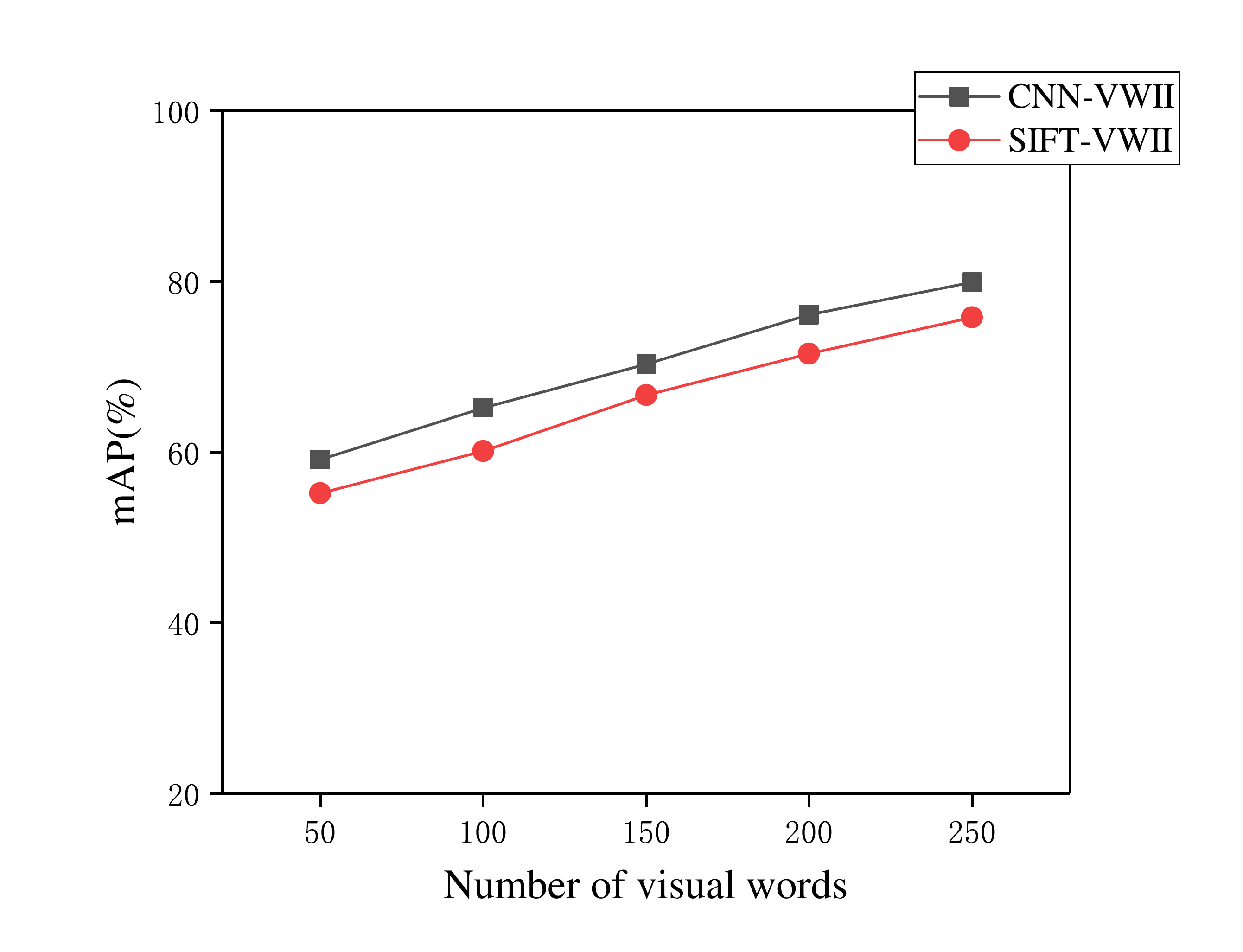}%
     }
     \subfigure[\footnotesize{Sports-1M}]{
     \includegraphics[width=0.48\linewidth]{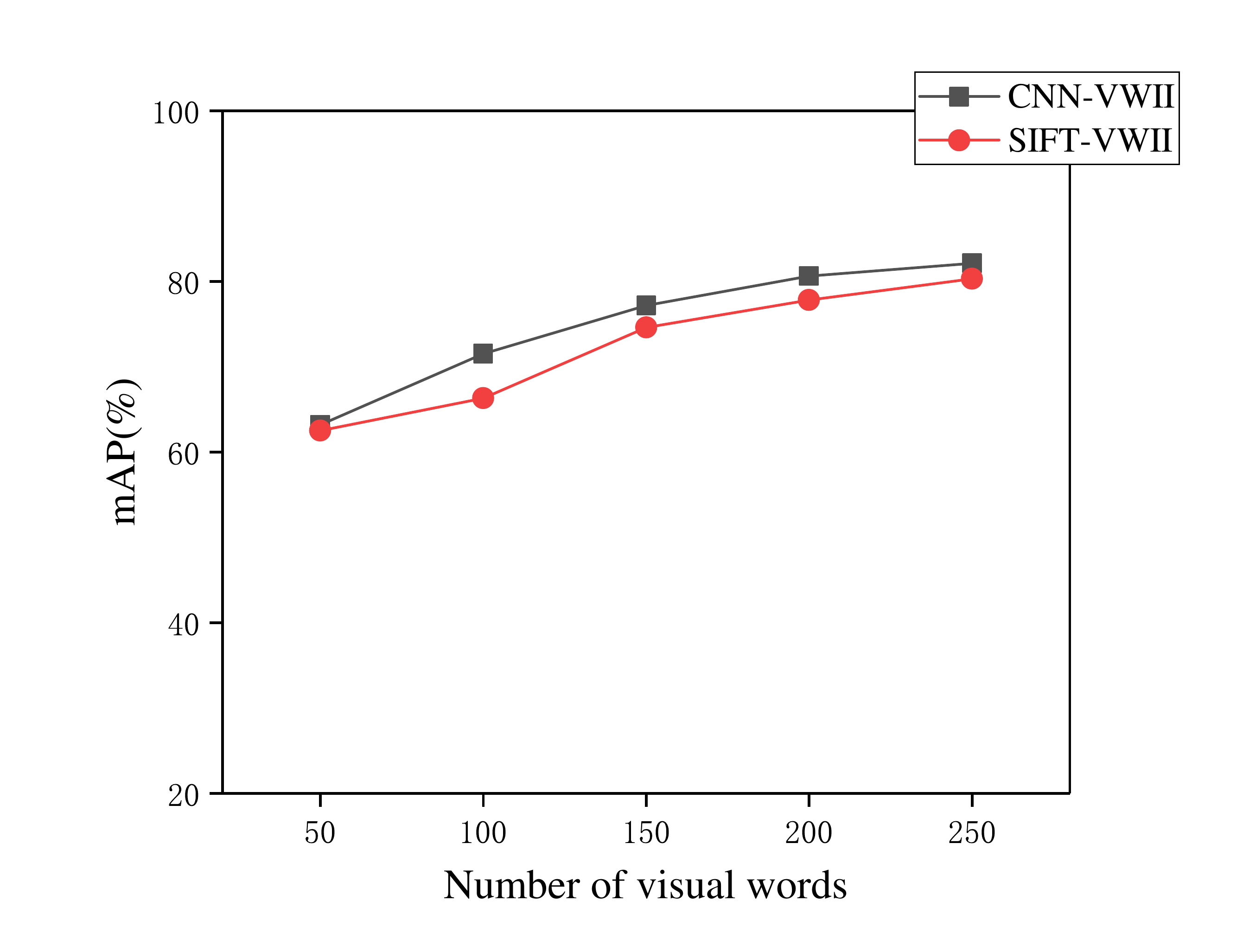}%
     }
     \captionsetup{justification=centering,font={footnotesize,sf}}
       \vspace{-0.3cm}
\caption{Precision evaluation on the number of visual words}
\label{fig:map-number-of-visual-words}
\end{center}
\end{minipage}

\vspace{-0.4cm}
\label{fig:k}
\end{figure*}

\begin{figure*}
\newskip\subfigtoppskip \subfigtopskip = -0.1cm

\begin{minipage}[b]{0.49\linewidth}
\begin{center}
     \subfigure[\footnotesize{YouTube-8M}]{
     \includegraphics[width=0.48\linewidth]{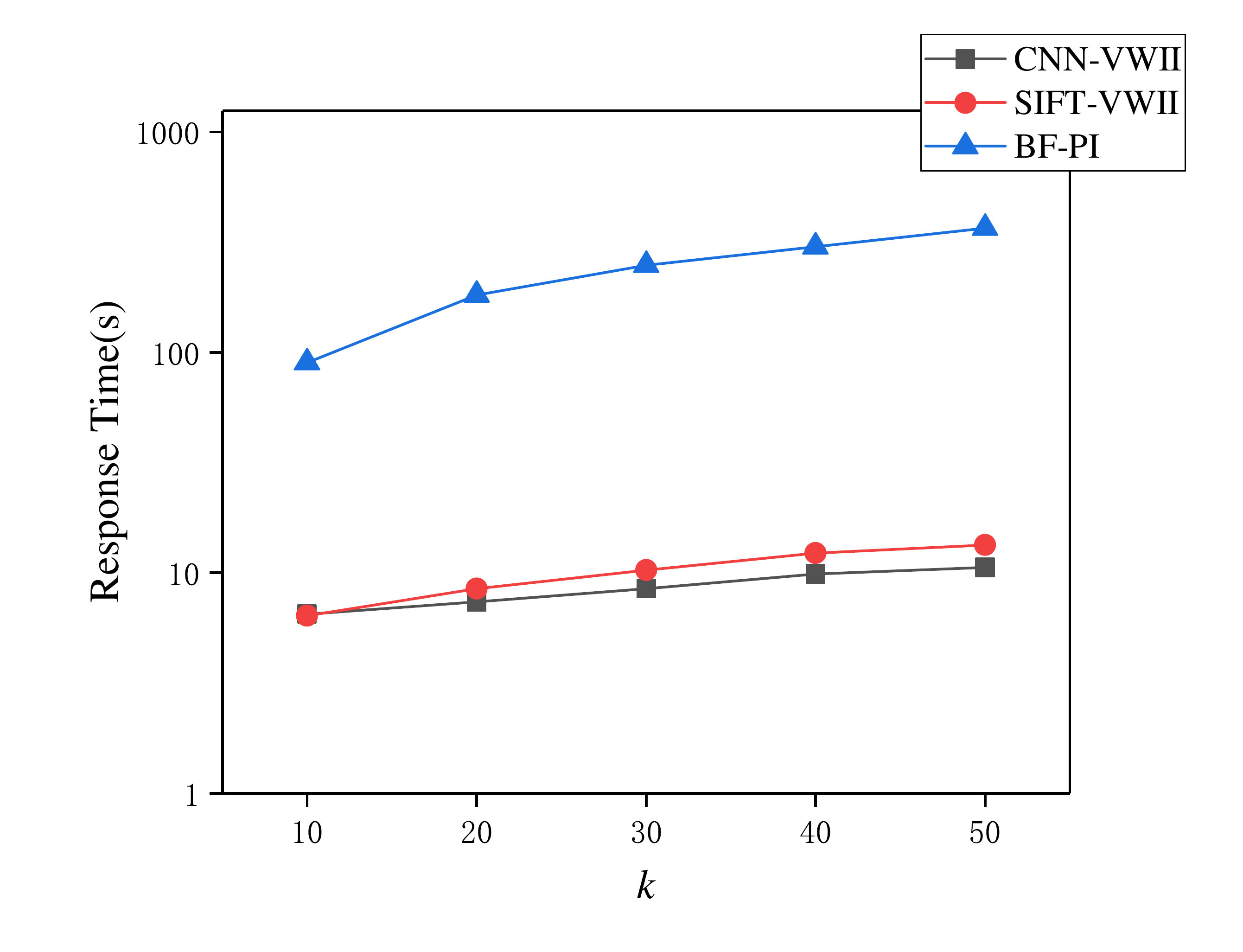}
     }
     \subfigure[\footnotesize{Sports-1M}]{
    \includegraphics[width=0.48\linewidth]{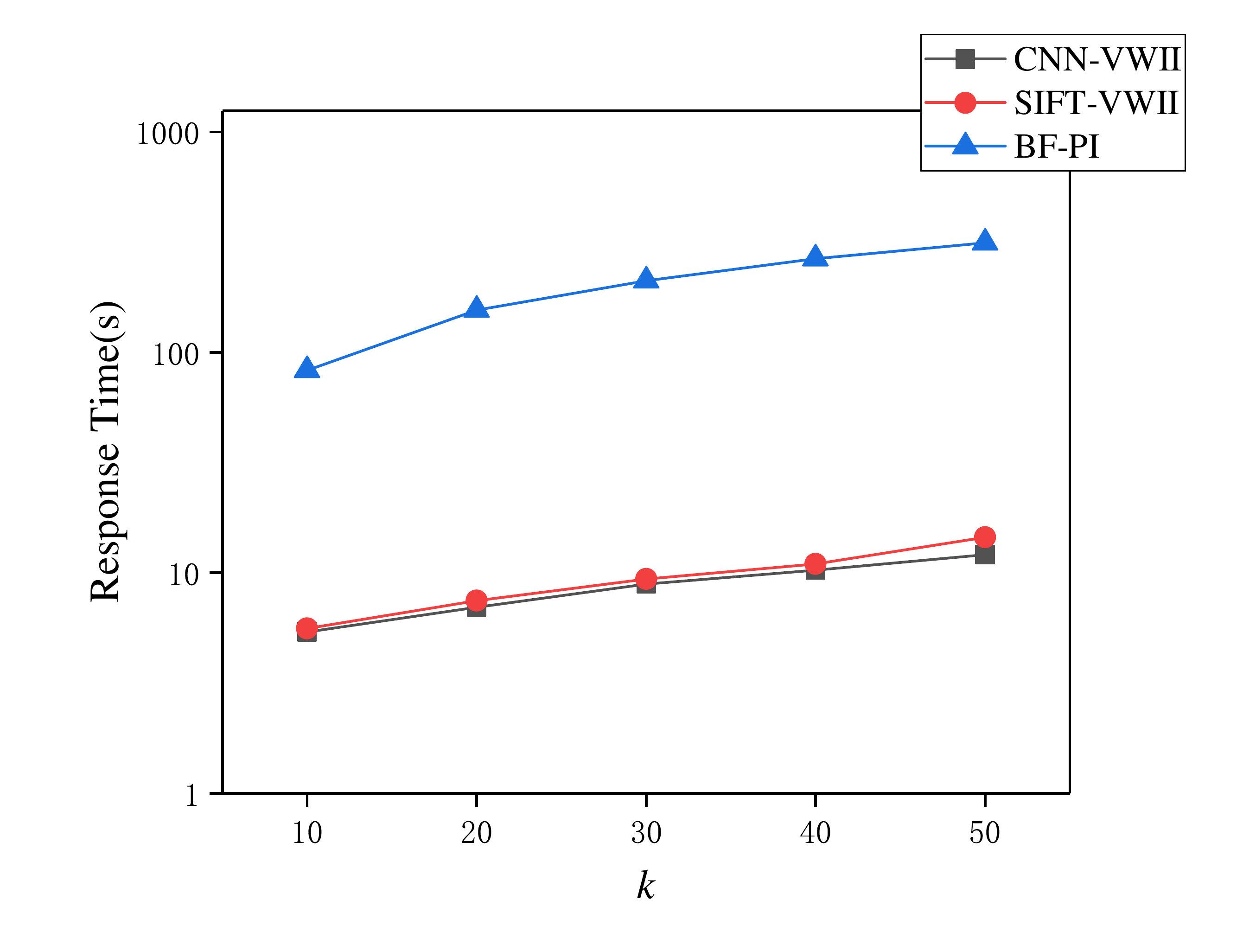}%
     }
     \captionsetup{justification=centering,font={footnotesize,sf}}
       \vspace{-0.3cm}
\caption{Precision evaluation on the number of results $k$}
\label{fig:response-time-k}
\end{center}
\end{minipage}
\begin{minipage}[b]{0.49\linewidth}
\begin{center}
     \subfigure[\footnotesize{YouTube-8M}]{
     \includegraphics[width=0.48\linewidth]{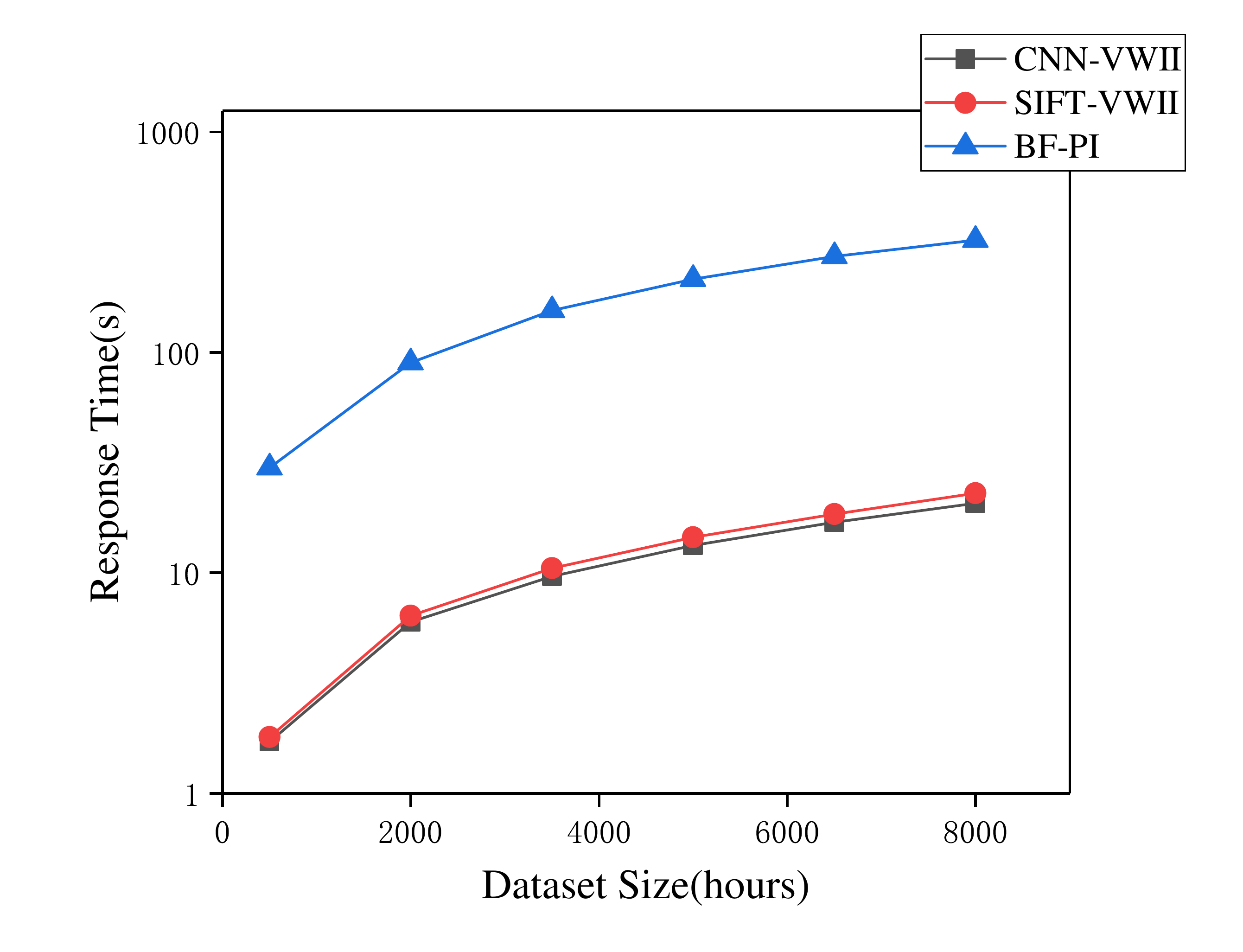}%
     }
     \subfigure[\footnotesize{Sports-1M}]{
     \includegraphics[width=0.48\linewidth]{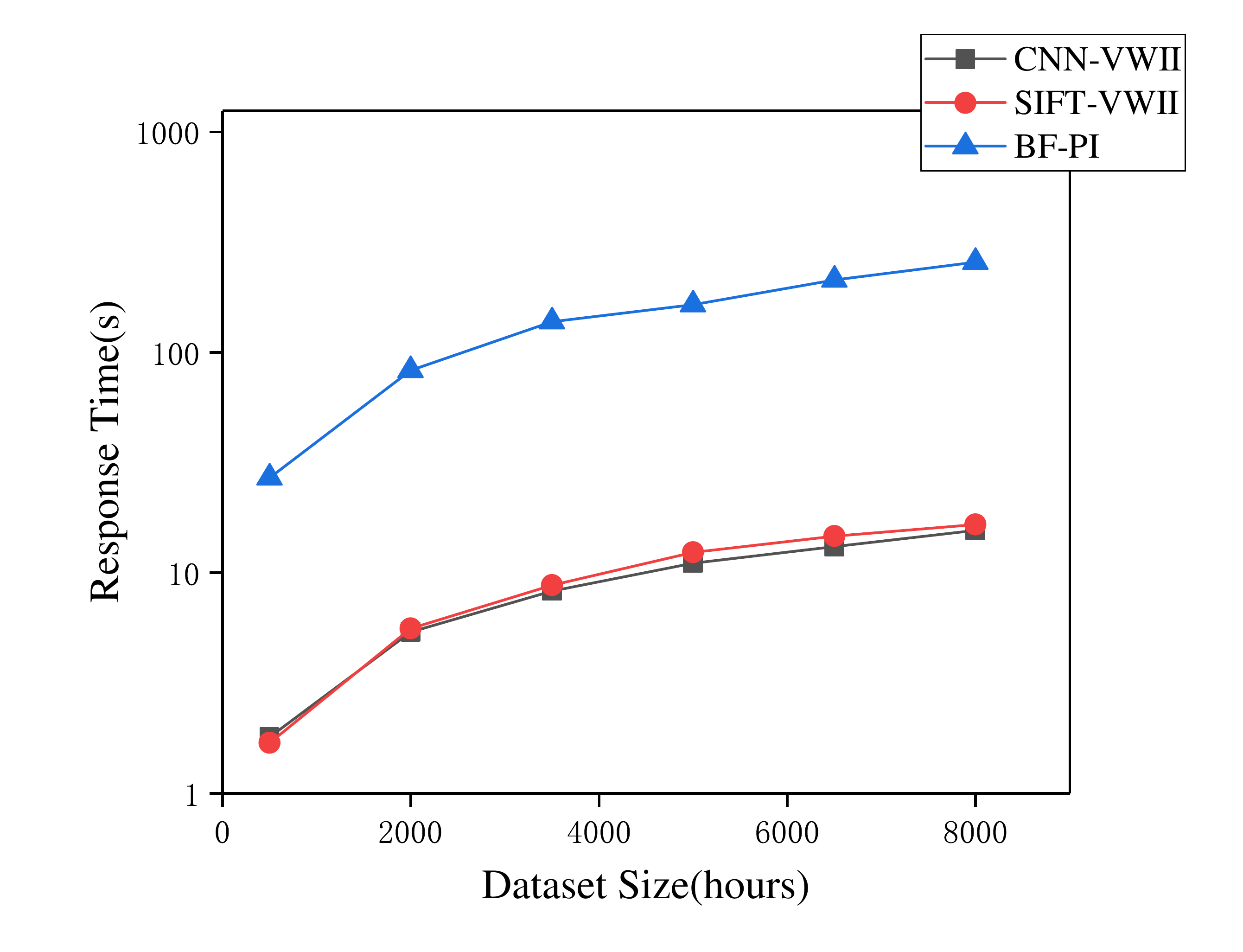}%
     }
     \captionsetup{justification=centering,font={footnotesize,sf}}
       \vspace{-0.3cm}
\caption{Efficiency evaluation on the size of dataset}
\label{fig:response-time-dataset-size}
\end{center}
\end{minipage}

\vspace{-0.4cm}
\label{fig:k}
\end{figure*}

In this section, we present results of a comprehensive performance study on real
video datasets to evaluate the accuracy, efficiency and scalability of the proposed techniques.
Specifically, we evaluate the accuracy and effectiveness of the following indexing techniques for top-$k$ image to video query.
\begin{itemize}
\item{\textbf{CNN-VWII}} CNN-VWII is the CNNs based visual weighted inverted index technique, which are proposed in our paper.
\item{\textbf{SIFT-VWII}} SIFT-VWII is the visual weighted inverted index technique, which is proposed in Section  ~\ref{index}, combines the traditional SIFT technique.
\item{\textbf{BF-PI}} BF-PI is the state-of-art technique, which is proposed in ~\cite{DBLP:journals/tcsv/AraujoG18}, designed for image to video retrieval.
\end{itemize}

\noindent\textbf{Dataset}. Performance of various algorithms is evaluated  comprehensively on two following video dataset. \textbf{YouTube-8M} (https://research.google.com/youtube8m/) is a large-scale labeled video dataset. It consists of 6.1 million you YouTube video IDs with high-quality machine-geneated annotations from a diverse vocabulary of more than 3800 visual entities. This dataset comes with precomputed audio-visual feature from 2.6 billions of frames and audio segments. Each video in it is between 120 aand 500 seconds long. \textbf{Sports-1M} (https://github.com/gtoderici/sports-1m-dataset) is another large-scale video dataset of 1 million YouTube videos belonging to a taxonomy of 487 classes of sports.

\noindent\textbf{Workload}. A workload for the top-$k$ image to video query consists of 100 queries. The query response time is employed to evaluate the performance of the algorithms. The video dataset size grows from 2000 to 8000 hours; the number of the query visual words changes from 50 to 150; the number of results $k$ changes from 10 to 50. By default, The video dataset size, the number of the query visual words, the number of results $k$ set to 2000, 100 and 10 respectively. Experiments are run on a PC with Intel Xeon 2.60GHz dual CPU and 16G memory running Ubuntu. All algorithms in the experiments are implemented in Java and Python.

\noindent\textbf{Precision evaluation on the size of dataset}. We evaluate the effect of the size of dataset on YouTube-8M and Sports-1M for precision evaluation. Fig.~\ref{fig:map-dataset-size}(a) illustrates that the precision of CNN-VWII, SIFT-VWII and BF-PI decrease with the rising the size of YouTube-8M. Specifically, when dataset size is 500 hours, the mAP of BF-PI is the highest among them. However, its precision drops faster than CNN-VWII in the interval of $[2000,8000]$. It is clearly that our method, CNN-VWII, has a better performance with the enlarging of the scale of data set. On Sports-1M dataset shown in Fig.~\ref{fig:map-dataset-size}(b), we can see that the mAP of all the three method descend step by step, similar to the situation on YouTube-8M. However, the precision of CNN-VWII is the highest all the time. Besides, when the dataset size if 500, the mAP of SIFT-VWII is a litter higher than BF-PI. In the interval of $[3500,8000]$, the precision of these two methods show a downward and fluctuating trend, lower than our method.

\noindent\textbf{Precision evaluation on the number of visual words}. We evaluate the effect of the number of visual words on YouTube-8M and Sports-1M for precision evaluation. As the method BF-PI do not generate visual words, we just compare the results of CNN-VWII and SIFT-VWII. In Fig.~\ref{fig:map-number-of-visual-words}(a), it is no doubt that with the increasing of the visual words number, both of them grow gradually. the mAP of our method is higher than SIFT-VWII all the time as the advantage of CNN on visual representation. The situation on Sports-1M dataset is a little different from YouTube-8M. In specific, shown in Fig.~\ref{fig:map-number-of-visual-words}(b), when the number of visual words increase from 100 to 150, the precision of SIFT-VWII rise faster and after that, the growth rate slows down. When the number of visual words is 250, the mAP of SIFT-VWII is near to CNN-SIFT.

\noindent\textbf{Efficiency evaluation on the number of results $k$}. Fig.~\ref{fig:response-time-k} demonstrates the results of evaluation on the number of results $k$ on YouTube-8M and Sports-1M for precision evaluation. The experiment results on YouTube-8M are shown in Fig.~\ref{fig:response-time-k}. With the increasing of $k$, the performance of BF-PI rises gradually, which is much higher than CNN-VWII and SIFT-VWII. The performance of our method is better than SIFT-VWII when $k \geq 20$ since the advantage of CNN in visual feature extraction and representation, which makes it easier to find the better results. On Sports-1M, illustrated in Fig.~\ref{fig:response-time-k}(b) CNN-VWII has the best performance among them, little higher than SIFT-VWII. No doubtly, the response time of BF-PI is the highest.

\noindent\textbf{Efficiency evaluation on the size of dataset}. We evaluate the effect of the size of dataset on YouTube-8M and Sports-1M for search efficiency. Not surprisingly, Fig.~\ref{fig:response-time-dataset-size}(a) shows that the response time of CNN-VWII, SIFT-VWII and BF-PI go up gradually with the rising of dataset size. More concretely, when the daataset size increases to 2000, the rising speed of all these method are fast. But in the interval of $[4000,8000]$, their growth becomes a little more gentle. It is obvious that the efficiency of CNN-VWII and SIFT-VWII are much higher than BF-PI because the using of index VWII can greatly improve search efficiency. The trends of CNN-VWII and SIFT-VWII are very similar. Fig.~\ref{fig:response-time-dataset-size}(b) shows that the evaluation on Sports-1M. Like the situation on YouTube-8M, the performance of CNN-VWII and SIFE-VWII are much better than BF-PI.

%% file: Conclution.tex
\section{Conclution}
\label{conclution}
In this paper, we study the problem of top-$k$ video retrieval by a query image, which aims to return $k$ most relevant video for a large-scale video database by a query image. We define top-$k$ image to video query formally and present the visual similarity function. To improve the retrieval precision and efficiency, we propose a novel model based on CNN and BoVW techniques for image and video representation. Besides, to boost the search efficiency, we design a novel indexing structure called VWII which is a combination of visual words and inverted index. Based on it we propose a novel algorithm for the task of top-$k$ image to video query. The experimental evaluation on two video datasets shows that our approach outperforms the state-of-the-art method.

%% file: Acknowledgement.tex
\section*{Acknowledgments}
This work was supported in part by the National Natural Science Foundation of China(61702560), the Natural Science Foundation of Hunan Province (2018JJ3691, 2016JC2011).